\documentclass[final,numberedheadings]{aipproc}
\usepackage{epsfig,amsmath,amssymb,mathrsfs,bm}
\usepackage[active]{srcltx}
\layoutstyle{6x9} 
\def\figdir{./}
\renewcommand\leq{\leqslant}\renewcommand\geq{\geqslant}
\def\vec#1{{\mathbf{#1}}}
\def\<{\langle}\def\>{\rangle}
\def\vn{\vec{n}}\def\vm{\vec{m}}\def\vk{\vec{k}}\def\vl{\vec{l}}
\def\vpsi{\bm{\psi}}
\def\MA{\mathbf{A}}\def\MB{\mathbf{B}}\def\MU{\mathbf{U}}
\def\MH{\mathbf{H}}

\def\wp#1{\widehat\partial_{#1}}

\def\c{\operatorname{c}}\def\s{\operatorname{s}}

\def\tee{\,\mathfrak{t}\,}\def\ell{\,\mathfrak{a}\,}
\def\emm{\,\mathfrak{m}\,}\def\ci{\,\mathfrak{c}\,}
\begin{document}
\title[]{Physics as Quantum Information Processing:\\ Quantum Fields as Quantum Automata \footnote{Work
    presented at the conference {\em , Foundations of Probability and Physics-6 (FPP6)} held on
    12-15 June 2011 at the Linnaeus University, V\"axj\"o, Sweden.}} 
 \author{Giacomo Mauro D'Ariano}{address={{\em QUIT} Group,
    Dipartimento di Fisica ``A. Volta'', 27100 Pavia,
    Italy, {\em http://www.qubit.it}\\
    Istituto Nazionale di Fisica Nucleare, Gruppo IV, Sezione di Pavia}}
\begin{abstract} 
  Can we reduce Quantum Field Theory (QFT) to a quantum computation? Can physics be simulated by a
  quantum computer? Do we believe that a quantum field is ultimately made of a numerable set of
  quantum systems that are unitarily interacting? A positive answer to these questions corresponds
  to substituting QFT with a theory of quantum cellular automata (QCA), and the present work is
  examining this hypothesis. These investigations are part of a large research program on a {\em
    quantum-digitalization} of physics, with Quantum Theory as a special theory of information, and
  Physics as emergent from the same quantum-information processing. A QCA-based QFT has tremendous
  potential advantages compared to QFT, being quantum {\em ab-initio} and free from the problems
  plaguing QFT due to the continuum hypothesis. Here I will show how dynamics emerges from the
  quantum processing, how the QCA can reproduce the Dirac-field phenomenology at large scales, and
  the kind of departures from QFT that that should be expected at a Planck-scale discreteness. I
  will introduce the notions of linear field quantum automaton and local-matrix quantum automaton,
  in terms of which I will provide the solution to the Feynman's problem about the possibility of
  simulating a Fermi field with a quantum computer.
\end{abstract}
\maketitle
\section{Introduction}
Quantum Theory (QT) is a theory of information: this has been shown in Ref. \cite{cdpax}, where QT
is derived from six fundamental assumptions about how information is processed.  Ref. \cite{cdpax}
achieves the goal of a long-term program initiated in 2003, which has been the topic of all my talks
at the V\"{a}xj\"o conferences on quantum foundations, since my first proposed set of axioms for QT
presented in 2005 \cite{my2005} (see Refs. \cite{myCUP2009,first} for an historical excursus).  The
program is certainly not finished, and one of the next tasks is to find other equivalent sets of
axioms.  One fact, however, is definitely established: QT is a theory of information, and whatever
set of axioms we consider, they must be equivalent to a set of rules for information processing.

The new informational point of view has far-reaching consequences on our way of understanding
physics, and has already shown its full power in greatly simplifying numerous derivations of
theoretical results of quantum information theory, reducing some theorems to tautologies. As for any
new really fundamental idea, the possible logical consequences and the new reasonings based on the
new principles are virtually unbounded, and many new interesting results are expected.

The six principles for QT have nothing of ``mechanical'' nature, since the QT derived in Ref.
\cite{cdpax} is just the general abstract {\em theory of systems}---Hilbert spaces, algebra of
observables, states, operations, and effects---and has no bearing on the {\em theory of
  mechanics}---particles, dynamics, and quantization rules: this is why I deliberately used the
words {\em Quantum Theory} in contrast to {\em Quantum Mechanics}.  However, Quantum Mechanics is
only a facet of the many sided Quantum Field Theory (QFT)---precisely its restriction to a fixed
number of particles---and QFT, strictly speaking, is itself a theory of systems.  It can be objected
that also QFT needs quantization rules, but many would agree that we should avoid them in a
autonomous foundation of QFT: on the contrary, it is classical mechanics the theory that should be
derived from QFT with a {\em classicalization} procedure.

But, how can we formulate a QFT autonomous from classical theory? The first thing to do is to take
QT as ``The Theory'' {\em ab initio}. The motivation is simple: QT as a theory of systems is solid
and autonomous. The idea is then to take QFT as the {\em theory of infinitely many quantum systems}.
Somebody may object that at least {\em space} and {\em time} are not systems, and would be missing
in such a theory. And what about Relativity?  The answer to this objection is that space and time
simply must be taken as non fundamental, but as derived notions: metric should emerge from the
purely discrete geometry of a network of quantum systems in interaction. With metric emerging from
pure topology, the only two additional principles will be: {\em topological homogeneity} and
{\em locality}---the informational equivalent of the universality of the physical law. The Minkowski
metric must follow only from locality and causality, the latter being the first of the six
informational principles for QT \cite{cdpax}.

In the above scenario QFT becomes a theory of the quantum cellular automata (QCA) that are local,
translational invariant, and reversible. This digitalization of QFT is the natural settings for the
holographic principle, with the QCA-based theory becoming a more fundamental layer for QFT, which in
turn should be regarded as only an effective theory, and we will then interpret the space-period of
the automaton $\ell$ to be the Planck length. One of the good news about a QCA-based QFT is that,
being naturally a lattice theory, it will also have the great bonus of avoiding all problems due to
the continuum which plague QFT (ultraviolet divergences, the Feynman path integral, non
localizability of measurements, and many more). The bad news is that the digital theory will likely
miss some of the simplicity of the continuum, and we must seek simple rules for interpolating the
digital with the analog descriptions, in order to make easy predictions of first-order corrections
to QFT and breakdown of covariance at the Planck scale, and being able to make comparison e.~g. with
doubly special relativity \cite{amelino}.

One striking feature of the computational paradigm is that large-scale Lorentz covariance is a free
bonus.  As a matter of fact, Lorentz covariance must emerge from the computation if the QCA is
capable of simulating physics. And, as shown in Ref. \cite{later}, the Dirac equation (at least for
space dimension $D=1$) is just the equation describing the free flow of quantum information. In Ref.
\cite{first} I showed how space-time can emerge from a topologically homogeneous causal network (as
a quantum circuit) in (1+1) dimensions, by building up foliations over the network, the Lorentz
time-dilation corresponding to an increased number of leaves within the tic-tac of an Einsteinian
clock, and space-contraction resulting from the inversely decreased density of events per leaf. In
Ref.  \cite{tosini} with A.  Tosini we have shown how a digital coordinate system can be obtained by
the analogous of the Einstein's procedure over the causal network, with signals sent back-and-forth
to events from an observer clock. This work is concerned with pure causality, without further
specification of the nature of the causal relation. For space dimension $D>1$ one has the {\em Weyl
  tiling} problem \cite{weyl}, stating the impossibility of emergence of an isotropic metrical space
from a discrete geometry.\footnote{Recently it has been proved in general that the maximal speed of
  information flow attainable in a periodic causal network is necessarily non isotropic, since the
  set of points attainable in a given maximum number of steps is a polytope that does not approach a
  circle \cite{tobias}.} In Ref.  \cite{FQXi} I proposed that the same quantum nature of the
causality relation provides a way out, with superposition of paths restoring the isotropy of maximal
speed of propagation of information, and I will discuss this here shortly in the concluding section.

In Refs. \cite{later,vaxjo2010,QCM2010} I introduced the QCA that implements the Dirac equation in
(1+1) dimension.There are already simulations of the Dirac field in the literature, noticeably the
lattice-gas realization of Ref. \cite{yepez}, however, they are are not QCA implementations, and
work only in 1st quantization and are designed for taking the continuum limit. In the same Refs.
\cite{later,vaxjo2010,QCM2010} I showed that the euristical derivation of the Dirac equation as free
flow of information along the network also provides a kinematical definition of the inertial mass in
terms of the slowing down of the information propagation due to the zig-zag motion over the causal
network, whereas the Planck constant becomes just a conversion factor between the informational mass
and the customary mass. In the QCA realization of the Dirac equation unitarity in conjunction with
discreteness introduces a renormalization of the speed of light, corresponding to a refraction index
of vacuum that is monotonically decreasing vs mass, with complete halt of the information flow at
the Planck mass---a new phenomenology coming from the informational point of view.  I also showed
how Fermi fields can be eliminated, having a QCA working locally on qubits: for $D=1$ space
dimension this solves the problem raised by Feynman \cite{Feynman}, who questioned the possibility
of simulating a Fermi field by a quantum computer with local interactions.\footnote{Feynman says:
  {\em The question is, if we wrote a Hamiltonian which involved only these operators, locally
    coupled to corresponding operators on the other space-time points, could we imitate every
    quantum mechanical system which is discrete and has a finite number of degrees of freedom? I
    know, almost certainly, that we could do that for any quantum mechanical system which involves
    Bose particles. I'm not sure whether Fermi particles could be described by such a system. So I
    leave that open. Well, that's an example of what I meant by a general quantum mechanical
    simulator.  I'm not sure that it's sufficient, because I'm not sure that it takes care of Fermi
    particles.}}  Here in Sect.  \ref{s:MQCA}, after reviewing this solution for $D=1$, I will also
provide the solution for $D=2$, which also generalizes to $D=3$.

We have now a fully {\em ab-initio quantum} field theory made with a QCA. The classical field
Hamiltonian is derived from the QCA as emergent, and this also opens the route to the mentioned
classicalization procedure. In order to address the Feynman problem in precise terms I will also
introduce the convenient notions of {\em linear quantum field cellular automaton} (LQCA) and of {\em
  local-matrix quantum cellular automaton} (MQCA). I will prove that the vacuum of the MQCA is
unique and factorized into single qubits states for $D=1$, whereas for $D=2$ additional qubits are
needed corresponding to auxiliary Majorana fields, whose state is entangled.  I will also provide
analytical relations between the ``emergent'' and the ``time-interpolating'' Hamiltonians, and
devote an entire section to the Quantum Random Walks of the LQCA, namely its first-quantization,
providing the first numerical evaluations for one and two particle states in the Planck regime.  I
will conclude the paper with some open problems and a brief discussion about possible ways out.

\section{The Field-linear Quantum Cellular Automaton\label{Qinf}}
We will consider quantum fields on a one-dimensional lattice $\mathbb{Z}$. We will use the following
notations for the field:
\begin{enumerate}
\item $\phi_n$ ($n\in\mathbb{Z}$) denotes a generic scalar (Boson or Fermion) field;
\item $\bm\phi_n=\{\phi_n^\alpha\}$, $\alpha$ in a finite set, denotes the vector field whose component
  generate the local algebra of the automaton at $n$; 
\item for the specific case of Dirac field the letter $\phi$ is substituted by the letter $\psi$; 
\item $\varphi_n$ denotes a generic scalar anticommuting field.
\end{enumerate}
Later on we will also consider fields on a $D$-dimensional lattice with $D>1$, e.~g. $\mathbb{Z}^D$,
and will use the boldface notation $\vn$ for the labeling on the lattice. We consider the special
case of quantum cellular automaton, whose algebra evolution is assigned by a linear evolution of a
quantum field on the lattice.\footnote{Considering boundary conditions on a bounded lattice does not
    affect derivations as long as we contemplate evolutions for finite numbers of time-steps of Fock
  states, namely states localized over a quiescent ``vacuum'' (see the following), and we can take
  the evolution as formally unitary, even for unbounded lattice.} We now focus on the case of $D=1$,
and consider $D>1$ in Subsect. \ref{sd2Feynm} and in the concluding section.

In a linear quantum field cellular automaton (LQCA), the (formally unitary) operator $U$ of the evolution
transforms the field as follows
\begin{equation}
\bm\phi(t+\tee)=U^\dag\bm\phi(t) U=\MU\bm\phi(t),
\end{equation}
where $t$ is a multiple of the time $\tee$ of each step of the automaton. Since the evolution must
preserve the (anti)commutation relations for the field, the unbounded matrix $\MU:= \|U_{ij}\|$ must
be itself unitary. The inverse evolution thus is $ \bm\phi(t)=U\bm\phi(t+\tee) U^\dag=\MU^\dag
\bm\phi(t+\tee)$.

\begin{figure}[t]
\includegraphics[width=.5\textwidth]{\figdir 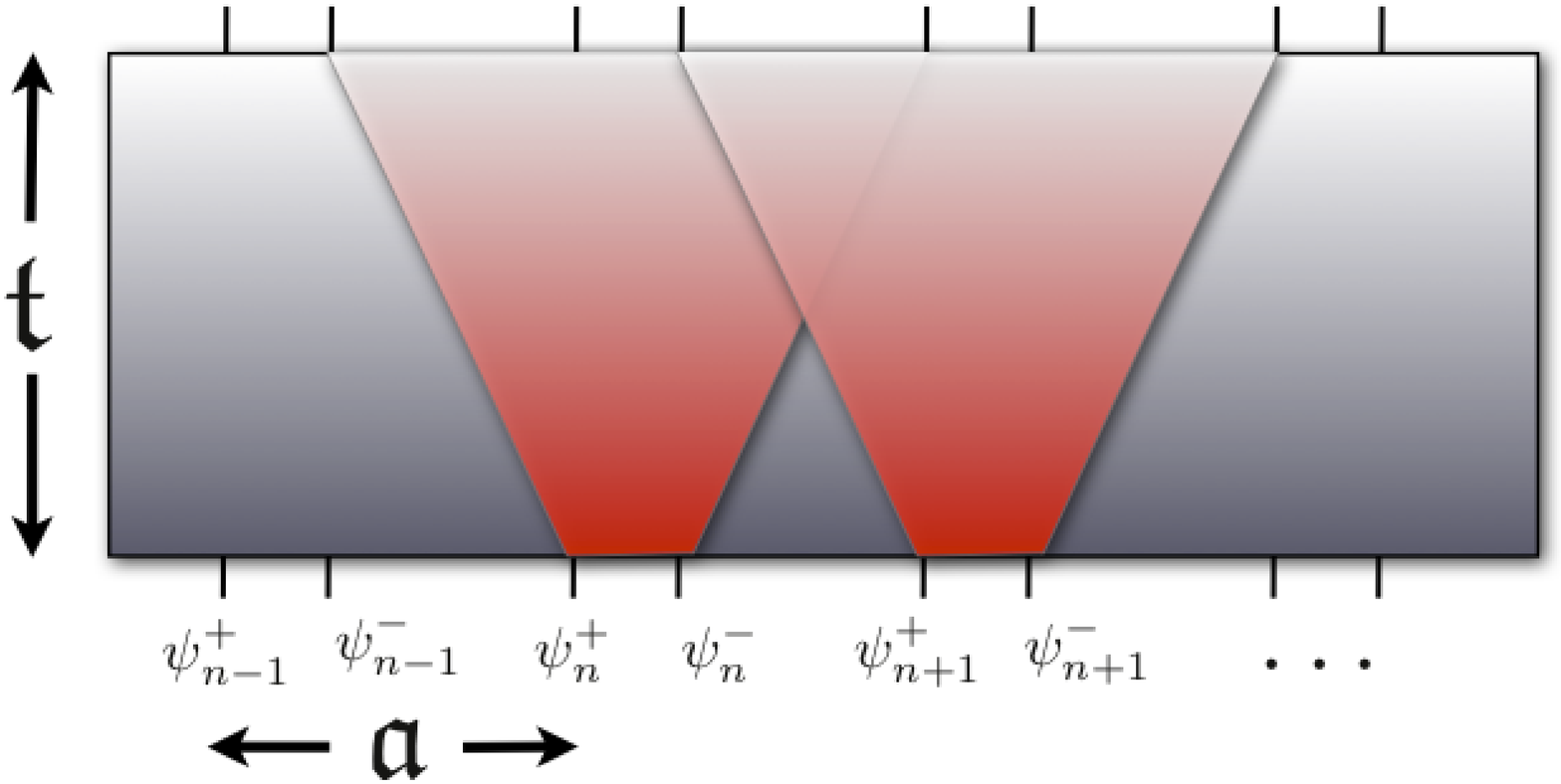}
\includegraphics[width=.5\textwidth]{\figdir 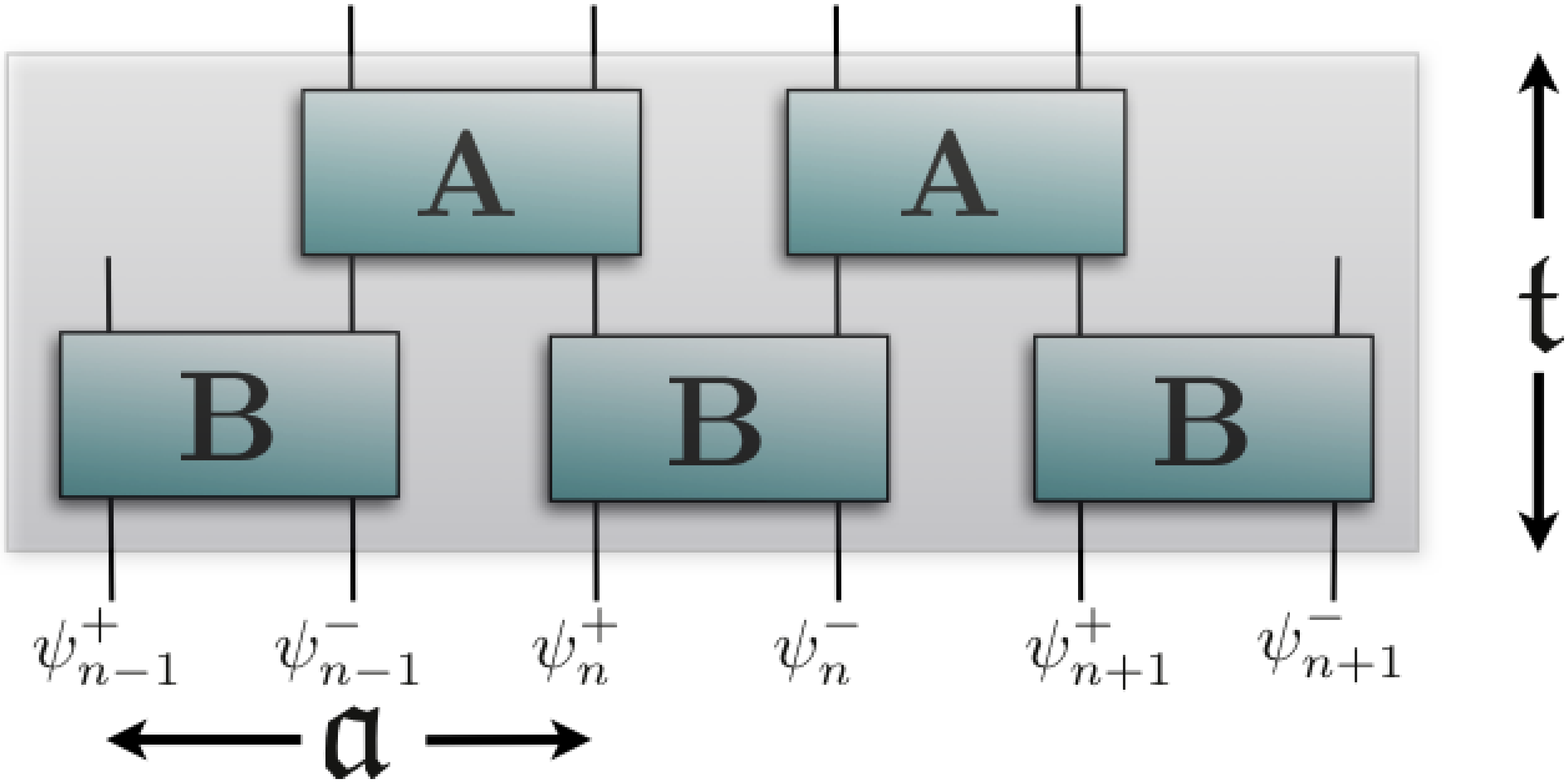}
\caption{(Left) Illustration of the linear cellular automaton for simulating the Dirac-Weyl equation
  in one space dimension. (Right) Margolus scheme for the automaton, namely the quantum circuit
  mimicking the Dirac equation.}
\label{f:automata}
\end{figure}

A LQCA evolves the operator algebra locally. This corresponds to having only a finite number of
nonvanishing elements in the rows and columns of the matrix $\MU$, namely $\MU$ is a {\em band
  matrix}. An example of LQCA is depicted in Fig. \ref{f:automata} with local algebra generated by
the Fermi-field vector
\begin{equation}\label{antipsi}
\vpsi:=\begin{bmatrix}\ldots\\\vpsi_n\\\vpsi_{n+1}\\\ldots\end{bmatrix},\qquad
\vpsi_n:=\begin{bmatrix}\psi_n^+\\\psi_n^-\end{bmatrix},\qquad
[\psi_n^\alpha,\psi_m^\beta{}^\dag]_+=\delta_{\alpha\beta}\delta_{nm}. 
\end{equation}
As proven in Ref. \cite{werner} a cellular automaton can be always decomposed into a finite number of
rows of independent gates connecting field operators in neighboring locations: this is the Margolus scheme
\cite{Margolus} (see Fig. \ref{f:automata}).
The cellular automaton is thus described by a quantum computation that is homogeneous both in space
and time, corresponding to a causal network with square lattice. On such lattice the Feynman 
path-integral is well defined, and corresponds to composing the evolution from many gates in
deriving the input-output relation of fields. The {\em path-sum rules} for the evolution have been
given in Ref. \cite{later}.

\section{The Dirac LQCA}\label{s:LQCA}
\begin{figure}[b]\includegraphics[width=.4\textwidth]{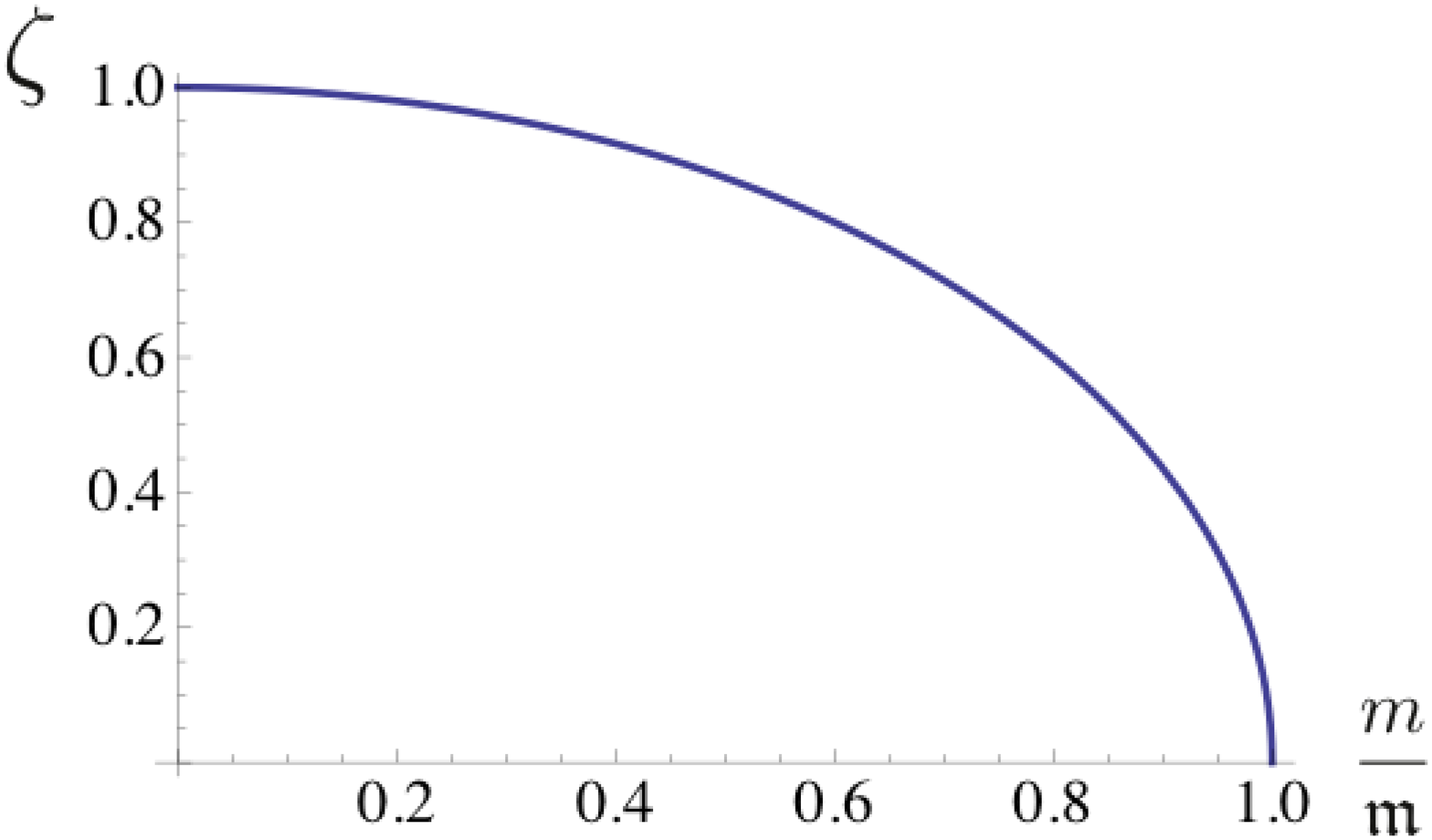}
  \caption{The inverse vacuum refraction index $\zeta$ versus the mass $m$ of the Dirac field. The
    mass scale is given by the Planck mass $\emm=\tfrac{\hbar}{\ell\ci}$, $\ell$ denoting the period of the
    automaton (from Ref. \cite{later}). \label{f:plotbound}}
\end{figure}

The (spinless) Dirac equation in 1 space dimension in the Weyl representation is 
\begin{equation}\label{Weyl}
\partial_t\vpsi=(-\ci\sigma^3\partial_x-i\sigma^1\omega)\vpsi,
\end{equation}
where $\omega=\ci\lambda^{-1}$, $\lambda=\hbar/m\ci$ is the Compton wavelength, and with vectors
$|\uparrow\>$ and $|\downarrow\>$ representing right-handed and left-handed fields, respectively,
whereas particle and antiparticle fields correspond to vectors $|+\>$ and $|-\>$, respectively. In
matrix form we have
\begin{equation}\label{Diracpde}
\partial_t\begin{bmatrix}\psi^+\\\psi^-\end{bmatrix}=
\begin{bmatrix}-\ci\partial_x &- i\omega\\- i\omega &\ci\partial_x\end{bmatrix}
\begin{bmatrix}\psi^+\\\psi^-\end{bmatrix}.
\end{equation}
The field $\vpsi$ satisfies the anti-commutation relations
$[\psi^\alpha(x),\psi^\beta(y){}^\dag]_+=\delta_{\alpha\beta}\delta(x-y)$, $x,y\in\mathbb{R}$
denoting values of the space coordinate.  We now want to derive a LQCA corresponding to the
discretized version of the Dirac equation. We label the field $\vpsi_n$ with a discrete index
$n\in\mathbb{Z}$ corresponding to discrete positions $n\ell$, with $\ell$ the periodicity of the
LQCA. The LQCA version of the Weyl equation would be the same as Eq. (\ref{Weyl}) with the partial
derivatives substituted by finite differences (it is convenient to consider finite differences in
the symmetric form $\widehat\partial:=\tfrac{1}{2}(\widehat\partial_+-\widehat\partial_-)$, with
$\partial_\pm f(n):=f(n\pm 1)$). In the quantum automaton version we will take for the speed of
light $\ci$ the {\em causal speed} $\ci=\ell/\tee$, namely the maximal speed of propagation in the
automaton.  Then, due to the unitarity, a discretized version of the Dirac partial differential
equation (\ref{Diracpde}) is possible only if we allow for a renormalization $\ci\to\zeta\ci$
\cite{later} (see also the following), namely
\begin{equation}\label{myDirac}
\widehat\partial_t\begin{bmatrix}\psi_n^+\\\psi_n^-\end{bmatrix}=
\begin{bmatrix}-\zeta\ci\widehat\partial_x &- i\omega\\- i\omega &\zeta\ci\widehat\partial_x\end{bmatrix}
\begin{bmatrix}\psi_n^+\\\psi_n^-\end{bmatrix},
\end{equation}
where $\widehat\partial_t=\tee^{-1}\widehat\partial$ and
$\widehat\partial_x=\ell^{-1}\widehat\partial$ (with the shift $\widehat\partial_\pm$ in the
appropriate discrete variable), $\tee$ denoting the execution time of the LQCA. The time-difference
in the LQCA (\ref{myDirac}) corresponds to the difference between the unitary matrices $\MU$ and
$\MU^\dag$, given by
\begin{equation}\label{sc}
\tfrac{1}{2}(\MU-\MU^\dag)=\begin{bmatrix}-\s\wp{}&-i\c\\-i\c&\s\wp{}\end{bmatrix},\quad
\MU=\begin{bmatrix}\s\wp{-}& -i\c\\ -i\c&\s\wp{+}\end{bmatrix},\;
\MU^\dag=\begin{bmatrix}\s\wp{+}& +i\c\\ +i\c&\s\wp{-}\end{bmatrix}.
\end{equation}
where 
\begin{equation}\label{sczeta}
\c=\omega\tee=\ell/\lambda,\quad\s=\zeta,
\end{equation}
and with unitarity implying the identity
\begin{equation}
\c^2+\s^2=1.
\end{equation}
Using Eq. (\ref{sczeta}) and parameterizing $\c$ and $\s$ by an angle $\theta$, one has
\begin{equation}
\c=\cos\theta=\frac{\ell}{\lambda}=\frac{m}{\emm},\quad\s=\sin\theta=\zeta=
\sqrt{1-\left(\frac{m}{\emm}\right)^2},\quad\frak{m}:=\frac{\hbar}{\ell\ci}, 
\end{equation}
which shows that $\zeta^{-1}$ is a mass-dependent vacuum refraction index which is strictly greater
than 1 for nonzero mass, monotonically increasing versus $m$, and becoming infinite at $m=\frak{m}$.
For $\frak{a}$ the Planck length $\frak{m}$ is the Planck mass, and the automaton becomes stationary
(i.~e. there is no propagation of information) at the Planck mass: this interesting violation of
dispersion relation has been presented in Ref. \cite{later}, and is due only to discreteness in
conjunction with the unitarity of the automaton (see Fig. \ref{f:plotbound}).

\subsection{Margolus scheme for the Dirac LQCA}
The Dirac automaton (\ref{myDirac}) can be achieved with the Margolus scheme in Fig.
\ref{f:automata}. The form of the gates $A$ and $B$ can be derived assuming, without loss of generality,
unit determinants $|\MA|=|\MB|=1$. Then, one has \cite{later} $B_{11}=B_{22}=0$, $B_{12}=B_{21}=\pm
i$, $A_{21}=A_{12}=\mp i \s$, and $A_{22}=A_{11}=\mp\c$. In the following, we will adopt the solution
\begin{equation}\label{AB}
\MA=\begin{bmatrix}\c&i \s\\ i\s&\c\end{bmatrix},\quad\MB=\begin{bmatrix}0&-i \\ -i&0\end{bmatrix}.
\end{equation} 

\subsection{Emergent Hamiltonian for the LQCA}\label{semerg}
The LQCA has no Hamiltonian: all unitary transformations are far from the identity. The Hamiltonian,
becomes an ``emergent'' notion: it can be written in terms of the LQCA unitary matrix as follows
(we remind the vector notation in Eq. (\ref{antipsi}))
\begin{equation}\label{H}
H=\frac{i\hbar}{2\tee}\vpsi^\dag (\MU-\MU^\dag )\vpsi.
\end{equation}
It is easy to show that one has
\begin{equation}\label{Schr}
i\hbar\widehat\partial_t\vpsi=[\vpsi,H]
\end{equation}
and in this sense $H$ is an Hamiltonian associated to the LQCA. It is formally identical to the
classical field Hamiltonian, which gives the field equation via Poisson brackets.  Notice that Eq.
(\ref{H}) would also hold using the non Hermitian Hamiltonian $H=i\hbar\tee^{-1}\vpsi^\dag
(\MU_1-\MU_2^\dag )\vpsi$ satisfying Eq.  (\ref{Schr}) for a halved-time finite-difference
derivative, with $\MU_1$ and $\MU_2$ the unitary transformations associated to the two rows of gates
in the Margolus scheme.  Eq. (\ref{Schr}) provides a three-point automaton evolution-rule
$\vpsi(t+\tee)=\vpsi(t-\tee)-2i\tee\hbar^{-1}[\vpsi(t),H]$, which can be time-reversed as
$\vpsi(t-\tee)=\vpsi(t+\tee)+2i\tee\hbar^{-1}[\vpsi(t),H]$. Thus the automaton invertibility is due
to the existence of a three-point updating rule, and not to Hermiticity of the Hamiltonian: the fact
that Hamiltonian can be chosen Hermitian is a consequence only of time-homogeneity of evolution (the
association of reversibility of a cellular automaton with a three-time updating has been first
noticed in Ref.  \cite{Normbook}).

The mapping between Hamiltonian and unitary evolution operator for the LQCA can be considered the
discrete version of the operator exponential mapping. One has 
\begin{equation}
\MU=\exp[-i\sin^{-1}(\MH\tee/\hbar)]
\end{equation}
in terms of the matrix
\begin{equation}
\MH:=\frac{i\hbar}{2\tee}(\MU-\MU^\dag)=[[\vpsi,H],\vpsi^\dag]_+,
\end{equation}
corresponding to 
\begin{equation}
H=\vpsi^\dag\MH\vpsi.
\end{equation}
This Hamiltonian is different from the {\em time-interpolating Hamiltonian} $\tilde H$ defined
through the identity
\begin{equation}
U=:\exp(-i\tilde H\tee/\hbar),
\end{equation}
which instead is given by 
\begin{equation}
\tilde H=\vpsi^\dag(\tee/\hbar)^{-1}\sin^{-1} (\MH\tee/\hbar)\vpsi,
\end{equation}
and is generally highly non local.

\subsection{The Feynman problem: simulating Fermi fields with a 
quantum computer}\label{s:MQCA} 
The algebra generated by the field in a given point $\vec{n}$ of a lattice, e.~g.
$\vec{n}\in\mathbb{Z}^D$, is what is usually referred as ``local'' or ``quasi-local'' algebra in the
literature, and this is what we called local algebra in the definition of the field quantum cellular
automaton. For Fermi fields, however, this locality notion is artificial, since the field
anticommutes at different positions and is itself non observable, as opposed to the case of a Bose
field. As mentioned in the introduction, the possibility of simulating a Fermi field by a quantum
computer---shortly {\em qubit-izing} Fermi fields---is still an open a problem, and was raised by
Feynman thirty years ago \cite{Feynman}. The solution to this problem is very relevant, since it
would respond also to the general question (also raised by Feynman in the same Ref. \cite{Feynman})
{\em whether Physics can be simulated by a universal (quantum) computer}: this is also the statement
of the Deutsch's version of the Church-Turing principle \cite{Deutsch}.

Let's state the Feynman problem more precisely. Since the case of Bose field is obvious (commutation
is trivially achieved by tensor product of local algebras), I state the problem only for the Fermi
case. We will call a quantum cellular automaton with algebra given by the tensor product of local
algebras that are finite-dimensional (with dimension independent on the size of the automaton in the
bounded case) {\em local-matrix QCA} (MQCA). We remind that the locality of the evolution is a
crucial part of the definition of QCA. With the above in mind, we now raise the fundamental problem
\begin{itemize}
\item[] {\bf Feynman problem:} Is it possible to
  find a local-matrix QCA for any Fermi LQCA?
\end{itemize}
For LQCA in $D=1$ space dimension a positive answer has been given in Ref. \cite{later}, which will
be shortly reviewed here. Thereafter I will also provide a solution for $D=2$ case, which can be
simply extended also to $D=3$.

The Fermi field can be put into one-to-one correspondence with an algebra generated by a tensor
product of local Pauli algebras by a relation of the form
\begin{equation}\label{idJW}
\sigma_\vec {n}^+=\varphi_\vec {n}^\dag \Phi(\vec{n}),
\end{equation}
where the ``phase factor'' $\Phi(\vec{n})\geq 0$ is a nonlocal functional of $\varphi_\vec{n}$.
Analytical expressions for $\Phi(\vec{n})$ has been given by Jordan and Wigner in Ref. \cite{JW} for
$D=1$, and for $D=2$ and $D=3$ by several authors in Refs.  \cite{Fradkin,Eliezer} and Ref.
\cite{Huerta}, respectively. Unfortunately, as we will see in the following, these expressions
cannot be used to solve the Feynman problem for $D>1$: to such purpose, a selfadjoint
$\Phi(\vec{n})$ is needed, and this new JW relation will be presented here.

\subsection{Solution of the Feynman problem for space dimension 1}\label{s:solu1}

Let me briefly review the solution of Ref. \cite{later}. The Jordan and Wigner \cite{JW} form of the
phase factor $\Phi(n)$ is given by
\begin{equation}\label{eJW}
\Phi(n)=\exp\left(i\pi\sum_{k<n}\varphi_k^\dag\varphi_k\right),
\end{equation}
and identity (\ref{idJW}) is inverted as follows
\begin{equation}
\varphi_j=\prod_{k<j}(-\sigma_k^3)\sigma_j^-.\\
\end{equation}
We see that the Pauli-local algebras are non-local in the field-local algebras, and viceversa the
field-local algebras are non-local in the Pauli-local algebras. 
Using the following identity that holds for both commuting and anticommuting $z,z'$
\begin{equation}
e^{\alpha^* z'^\dag z-\alpha z^\dag z'} z e^{\alpha z^\dag z'-\alpha^*z'^\dag z} =\cos|\alpha|\,z
+\frac{\alpha}{|\alpha|}\sin|\alpha|\,z',
\end{equation}
it is easy to check that 
\begin{equation}
\MA_n\begin{bmatrix}\psi^-_{n-1}\\ \psi^+_n\end{bmatrix}=
A_n^\dag\begin{bmatrix}\psi^-_{n-1}\\ \psi^+_n\end{bmatrix}A_n,\quad
\MB_n\begin{bmatrix}\psi^+_n\\ \psi^-_n\end{bmatrix}=
B_n^\dag\begin{bmatrix}\psi^+_n\\ \psi^-_n\end{bmatrix}B_n,
\end{equation}
with
\begin{equation}
A_n=\exp\left[i\theta\left(\psi_n^+{}^\dag\psi_{n-1}^-+\psi_{n-1}^-{}^\dag\psi_n^+\right)\right],\qquad
B_n=\exp\left[-\tfrac{i\pi}{2}\left(\psi_n^+{}^\dag\psi_n^-+\psi_n^-{}^\dag\psi_n^+\right)\right].
\end{equation}
Using the JW identity (\ref{eJW}) with $\psi^+_n=\varphi_{2n}$ and $\psi^-_n=\varphi_{2n+1}$,
the unitary operators of the gates rewrite as follows
\begin{equation}\label{ABU}
A_n=\exp\left[-i\theta\left(\sigma_{2n-1}^-\sigma_{2n}^++\sigma_{2n-1}^+\sigma_{2n}^-\right)\right],\quad
B_n=\exp\left[\tfrac{i\pi}{2}\left(\sigma_{2n}^+\sigma_{2n+1}^-+\sigma_{2n}^-\sigma_{2n+1}^+\right)\right].
\end{equation}
From the above construction it is also clear that any LQCA in $(1+1)$ dimension has a MQCA
realization, since it has a Margolus realization with two rows, and each gate unitary operator can
be expressed as exponential of bilinear form of the field in neighboring positions, say $n$ and
$n+k$, $\ldots$, which can be written as tensor products of local matrices in all positions starting
from $n$ and ending at $n+k$. Indeed, the JW construction leads to the ``string'' identity
\begin{equation}\label{string0}
\varphi_{j+l}^\dag\varphi_j=(-)^l\sigma_j^-S_j^l\sigma_{j+l}^+,\quad
S_j^l:=\prod_{j< k<j+l}\sigma_k^3,\qquad\text{(``string'' identity)},
\end{equation}
and only for neighboring locations ($l=1$) one has Pauli operators only at the same locations of the
fields, otherwise one has a string of additional $\sigma^3$ operators joining the two locations. Can
we reduce the number of qubits to those only at the same locations of the field? An idea for
answering to this question is provided by Ref.  \cite{cirvar}, and consists in introducing
additional Majorana fields $\varphi_j^1:=\varphi_j^\dag+\varphi_j$ and
$\varphi_j^2:=i(\varphi_j^\dag-\varphi_j)$ anticommuting with the original Dirac fields, and writing
both Dirac and Majorana fields in terms of a single JW string of Pauli operators, pertaining Dirac
and Majorana fields in alternate positions (in our case taking the even locations for the Dirac
and the odd ones for the Majorana). Then, one has the identity
\begin{equation}
\varphi_{j+2l}^\dag\varphi_j P_{j+2l+1,j+1}=-i\sigma_j^-\sigma_{j+2l}^+\sigma_{j+1}^2\sigma_{j+2l+1}^1,\quad
P_{ij}^\dag=P_{ij}:=i\varphi^1_i\varphi^2_j,\quad P_{ij}^2=I.
\end{equation}
The idea now is that if all $P_{ij}$ involved in the interaction terms
$\varphi_{j+2l}^\dag\varphi_j$ commute, then, upon jointly diagonalizing them on a common vacuum
state, one would have $\varphi_{j+2l}^\dag\varphi_j=\pm
i\sigma_j^-\sigma_{j+2l}^+\sigma_{j+1}^2\sigma_{j+2l+1}^1$, and the MQCA on that vacuum would be
local.  However, it is easy to check that $[P_{ij},P_{i'j'}]=[P_{ij},P_{ji}]=0$ for $i\neq i'$ and
$j\neq j'$, but $[P_{ij},P_{il}]_+=[P_{il},P_{jl}]_+=0$.  Therefore, the construction works if there
are interaction terms only at a single fixed distance $l$, not when there are more distances $l$ of
interactions.

\subsubsection{The Vacuum for the Dirac automaton}\label{s:vacuum}
There are many definitions of {\em vacuum} in QFT, and they are no longer equivalent in the case of
the quantum automata. In order to build up a Fock space for the LQCA, we need to require that the
vacuum $|\Omega\>$ be annihilated by the field operator $\varphi_n$, namely
\begin{equation}
\varphi_n|\Omega\>=0.
\end{equation}
Notice that in the present context such definition has no apparent relation to the vacuum definition
as the lowest-eigenvalue eigenstate of the Hamiltonian, even for the emergent Hamiltonian (\ref{H}),
since this is not jointly diagonal with $U$. On the other hand, as we will see in Sect.
\ref{s:1st}, $|\Omega\>$ is left invariant by the unitary evolution.

In the JW representation the only possible vacuum for the $D=1$ Dirac MQCA is
\begin{equation}\label{vacuum}
|\Omega\>=\otimes_k|\!\downarrow\>_k.
\end{equation}
{\bf Proof:} Write the vector $|\Omega\>$ as 
$|\Omega\>=:|\!\downarrow\>_n|\Omega_n^\downarrow\>+|\!\uparrow\>_n|\Omega_n^\uparrow\>$,
with $|\Omega_n^\uparrow\>$ and $|\Omega_n^\downarrow\>$ (generally are unnormalized and non
orthogonal) states for the algebra of all qubits at location $m\neq n$. Being $\Phi(n)$
an exponential, it cannot annihilate any vector, whence $\varphi_n|\Omega\>=|\!\downarrow\>_n
\Phi(n)|\Omega_n^\uparrow\>$, which is null if and only if $|\Omega_n^\uparrow\>=0$. Upon repeating
the same reasoning for all $n$, the statement (\ref{vacuum}) follows. $\blacksquare$ 

Notice the useful identity: $\varphi_n^\dag|\Omega\>=\sigma_n^+|\Omega\>$.

\subsubsection{Relation with one-dimensional models in statistical mechanics}
It interesting to notice that the Hamiltonian (\ref{H}) in the JW realization of $D=1$ Dirac field
corresponds to a spin model of the form
$H=-i\ci\zeta\ell^{-1}\sum_n\sigma^+_n\sigma^3_{n-1}\sigma^-_{n-2}-\omega\sum_n\sigma^+_{2n}\sigma^-_{2n-1}+
\operatorname{h.~c.}$. It would be interesting to see how an integrable spin model (solved e.~g. by
the Bethe ansatz and/or Yang-Baxter) is translated into a LQCA, in which case the LQCA would provide
an easy numerical simulation for the evolution for discretized time for first-quantization
situations as e.~g. spin-waves. Notice, however, that for example the conventional Hamiltonian for
the Heisenberg $XY$ model in transverse field corresponds to a Dirac conventional Hamiltonian
containing also a four-linear interaction term, indicating that the field-linearity of the automaton
may not be preserved in the correspondence.

Another interesting relation with condensed-matter theoretical models is the case of Hamiltonians of
the form $H=-\hbar\kappa\sum_j(c^\dag_{j+1}c_j+c^\dag_jc_{j+1})$, $\kappa$ a coupling constant,
describing noninteracting 
electrons hopping on a 1-dimensional lattice. There is noting relativistic about this: however, if
you fill the system up to the Fermi energy $\varepsilon_F$ and focus on electrons with energy
$E=\varepsilon-\varepsilon_F\ll \varepsilon_F$ and momentum $|k-k_F|\ll k_F$, the right and
left-moving fields associated with this electron will obey the zero-mass Weyl equation in 1 space
dimension. Then, a Peirls distortion would open an off-diagonal term in the energy spectrum
corresponding to a mass-coupling between the two fields (see e.~g. Ref. \cite{Zee}).

\subsection{Solution of the Feynman problem for space
 dimension D=2,3}\label{sd2Feynm}

In Ref. \cite{cirvar} it has been proposed to use Majorana auxiliary fields to eliminate fields in
Hamiltonian models for $D>1$. The idea is to use the construction mentioned in Subsect.
\ref{s:solu1}, based on covering the $D$-dimensional lattice with a 1-dimensional string.
Unfortunately, as I have shown, the construction works only if there are interaction terms at a
single given distance, whence it cannot work in the 1-dimensional covering, since there are always
more than two neighboring locations in interaction (the coordination number is necessarily greater
than 2). In addition, the 1-dimensional covering raises the problem that the bulk realization will
depend on boundary terms, and would be unsuitable for unbounded lattices. The idea of using Majorana
fields, however, remains valuable, and will be used here.  In constructing the Majorana fields, it
has been crucial to have the JW ``phase factor'' as formally selfadjoint $\Phi(n)=\Phi^\dag(n)$, so
that $\Phi(n)^2=I$, which is needed for the cancellation of strings of Pauli matrices.
Unfortunately the form proposedd in Refs.  \cite{Fradkin,Eliezer,Huerta} for $D=2,3$ do not satisfy
this requirement. We need then a new JW construction. 

In order to have identity (\ref{idJW}) connecting commuting Pauli matrices $\sigma_\vn$ with
anticommuting fields $\varphi_\vn$, one needs to satisfy the identity
\begin{equation}\label{phisigma}
\varphi_\vn\Phi(\vm)=\Phi(\vm)\varphi_\vn e^{i\alpha_\vn^{(\vm)}},
\end{equation}
with the phase $\alpha_\vm^{(\vn)}$ satisfying the identities 
\begin{equation}\label{alphas}
\alpha_\vm^{(\vn)}=\alpha_\vn^{(\vm)}+\pi,\quad \alpha_\vn^{(\vn)}=0.
\end{equation}
In order to satisfy identity (\ref{phisigma}) one needs to adopt the general form for the phase
factor
\begin{equation}
\Phi(\vn):=\exp\left(i\sum_\vm\varphi_\vm^\dag\varphi_\vm\alpha_\vm^{(\vn)}\right),
\end{equation}
whereas for $D=2$ identity (\ref{alphas}) along with the requirement that $\Phi$ is selfadjoint, imply the
following step-function analytical form for $\alpha_\vm^{(\vn)}$
\begin{equation}\label{thetafun}
\alpha_\vn^{(\vm)}=\theta_{2\pi}(\vm-\vn:\vec{i}),\quad \theta_{2\pi}(\vk)=
\begin{cases}0, & \text{if }0\leq\arg(\vk:\vec{i})<\pi,\\
\pi, & \text{if }\pi\leq\arg(\vk:\vec{i})<2\pi,\\
\end{cases}
\end{equation}
where $\arg(\vk:\vec{i})$ is the angle between the vector $\vk$ and the vector $\vec{i}$ directed as
the x-axis in the plane. It is immediate to check that $\Phi(\vn)^\dag=\Phi(\vn)$ and
$\Phi(\vn)^2=I$, and that for $D=1$ we recover the JW construction, whereas for $D=2$ the
construction corresponds to set the origin of the lattice in $\vn$ and put an identity in each site
of the upper half-plane (including the positive $x$ axis and the origin), and put a $\sigma^3$ on
all other sites. For $D=3$ we need to make a rotation by $\pi$ around the axis orthogonal to the
plane identified by vectors $\vn$ and $\vm$, and in order to keep the transformation of Dirac
spinors under rotations we can use the isospin form of the phase-factor suggested in Ref.
\cite{Huerta}, but with the angle again as a step-function (\ref{thetafun}), in order to satisfy the
additional requirement $\Phi(\vn)^\dag=\Phi(\vn)$. In the following we will focus for simplicity
only on the case $D=2$.

We can now add auxiliary Fermi fields e.~g. upon redefining the phase factor as follows
\begin{equation}
\Phi(\vn):=\exp\left[i\sum_\vm(\varphi_\vm^\dag\varphi_\vm+\vartheta_\vm^\dag\vartheta_\vm)
\alpha_\vm^{(\vn)}\right].
\end{equation}
and introduce corresponding Pauli matrices as follows
\begin{equation}
\tau_\vn^+=\vartheta_\vn^\dag\Phi(\vn).
\end{equation}
The auxiliary fields anti-commute with the original Fermi fields.\footnote{A. Tosini has shown that
  it is possible to make the Dirac and Majorana fields both commuting and anticommuting.} We can now
introduce the Majorana 
fields $\vartheta_\vn^\nu$, with $\nu=1,2$ corresponding to the Pauli matrices $\tau_\vn^\nu$ and
introduce the observables $P_{\vm\vn}:=i\vartheta_\vm^1\vartheta_\vn^2$.  It is easy to see that
$P_{\vn\vm}^2=I$, and for $\vn\neq\vn'$ and $\vm\neq\vm'$ one has
$[P_{\vn\vm},P_{\vn'\vm'}]=[P_{\vn\vm},P_{\vm\vn'}]=[P_{\vn\vm},P_{\vm\vn}]=0$,
$[P_{\vn\vm},P_{\vn\vm'}]_+=[P_{\vm\vn},P_{\vm'\vn}]_+=0$. Then one has the identity
\begin{equation}
\varphi_\vn ^\dag\varphi_\vm P_{\vn\vm}=-i\sigma_\vn^+\sigma_\vm^-\tau_\vn^1\tau_\vm^2.
\end{equation}
If we now orient all interaction links $\vl$ and take only $+\vl$ when we have both $\pm\vl$, then
$P_{\vn,\vn+\vl}$ for varying $\vn$ and $\vl$ commute each other (as long as we do not have
different $\vl$ that are parallel). One can then introduce a joint vacuum for both types of fields,
which jointly diagonalizes all the observables $P_{\vn,\vn+\vl}$. It is easy to see that in the
qubit representation such vacuum is factorized into separate components for $\sigma$ and $\tau$
algebras, and the derivation of Subsect. \ref{s:vacuum} still holds for the $\sigma$ part which is
then a tensor product of $|\downarrow\>$ whereas the $\tau$-vector is entangled. On such vacuum, one
has the identity
\begin{equation}
\varphi_\vn ^\dag\varphi_{\vn+\vl}+\varphi_{\vn+\vl}^\dag\varphi_\vn=\pm 
(\varphi_\vn ^\dag\varphi_{\vn+\vl}+\varphi_{\vn+\vl}^\dag\varphi_\vn)P_{\vn,\vn+\vl}
=\pm i(\sigma_\vn^+\sigma_{\vn+\vl}^--\sigma_{\vn+\vl}^+\sigma_\vn^-)\tau_\vn^1\tau_{\vn+\vl}^2.
\end{equation}
The same vacuum diagonalizing the observables $P_{\vn,\vn+\vl}$ does not diagonalize also
$\tau_\vn^1\tau_{\vn+\vl}^2$, and we need to work with a doubled qubits. 

\section{First quantization: Quantum Random Walks}\label{s:1st}
\begin{figure}[b]
\includegraphics[width=.35\textwidth]{\figdir 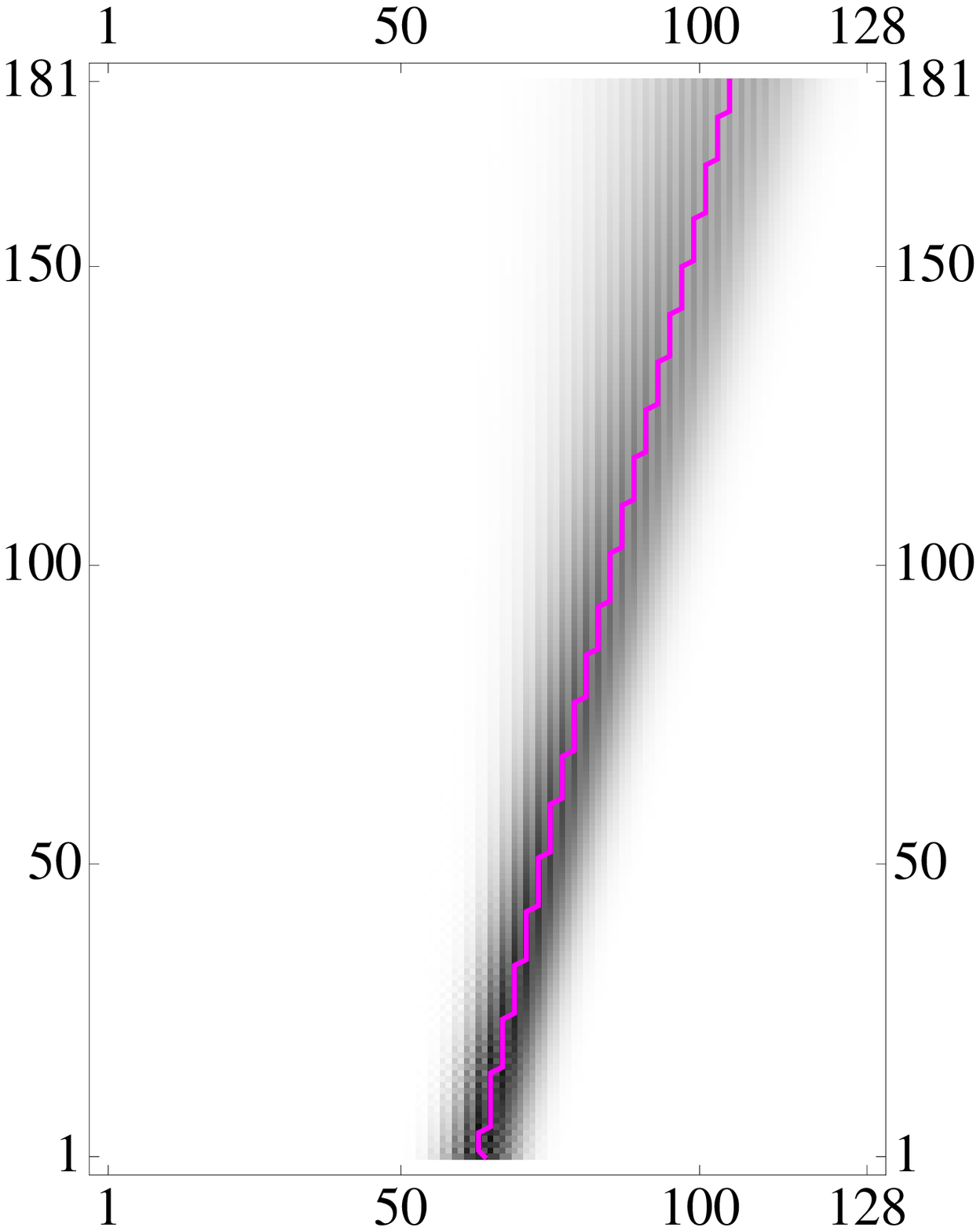} 
\includegraphics[width=.35\textwidth]{\figdir 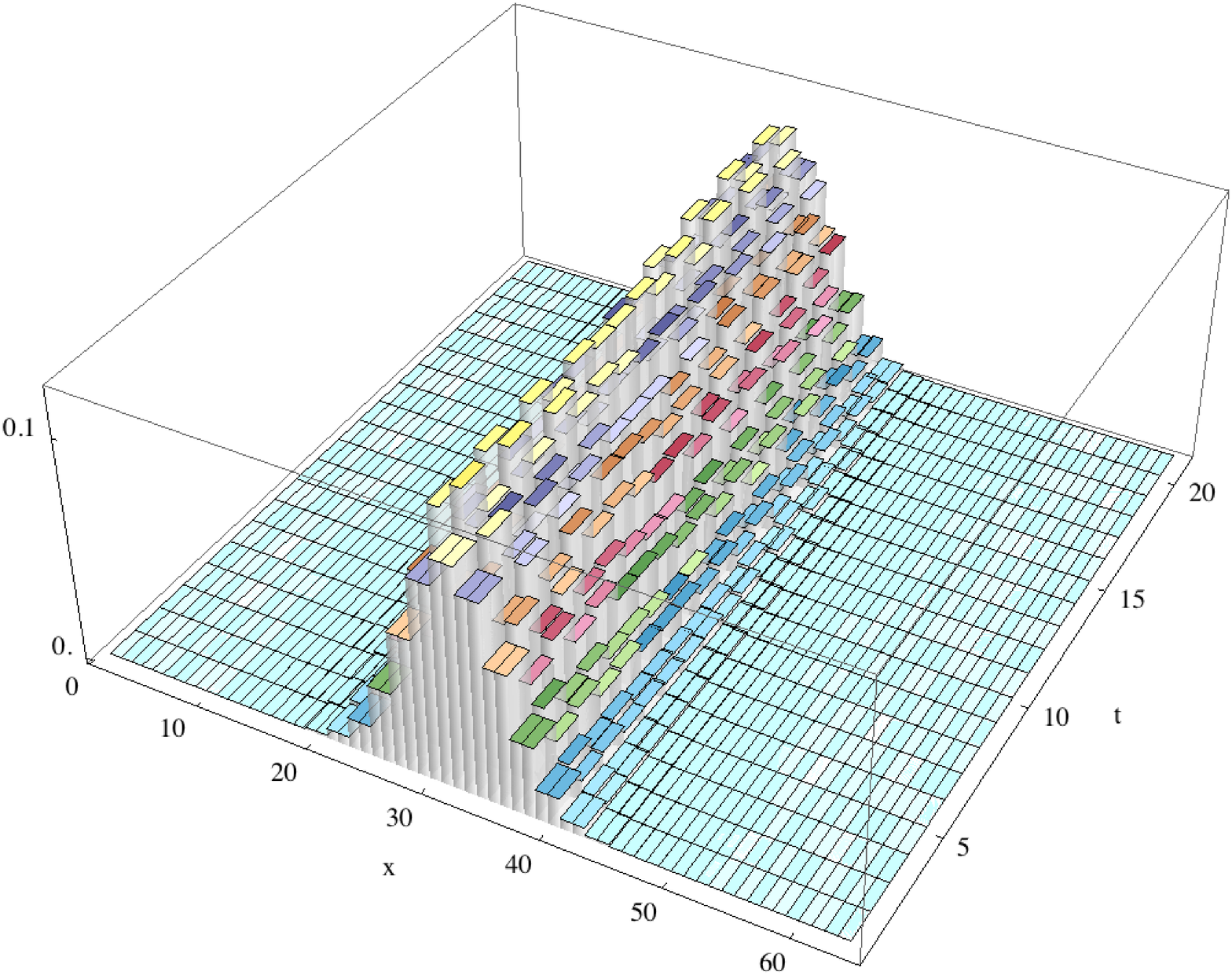} 
\includegraphics[width=.35\textwidth]{\figdir 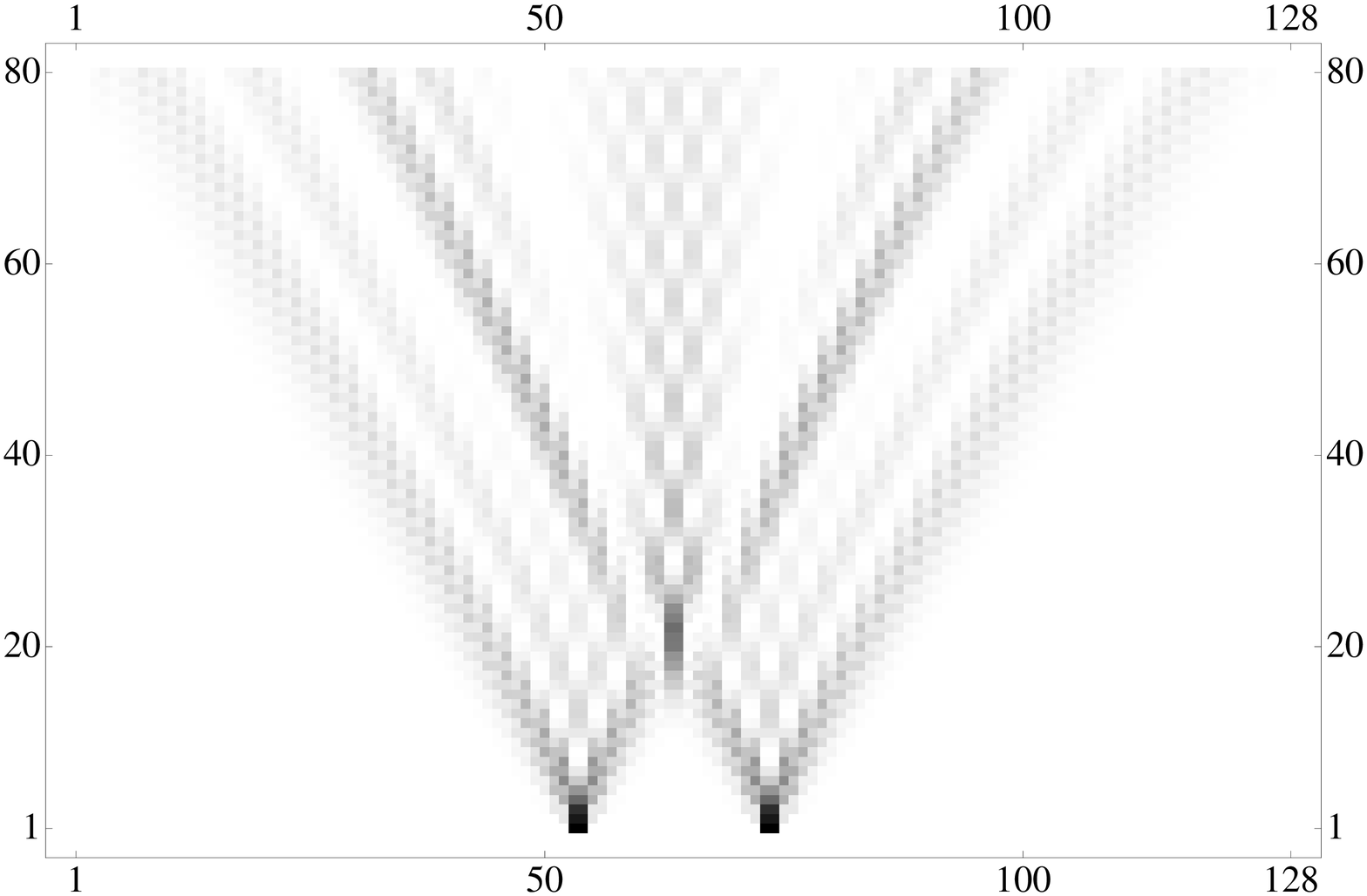} 
\caption{(Left) The evolution of a single-particle Gaussian packet of the form in Eq.
  (\ref{Gauss-pa}) with $x_0=0$, $\Delta=2$, $k=8$, for 180 time-steps (i.~e. $t=180\tee$) and a
  total dimension of $128$, corresponding to $128$ qubits, half of them for left and half of them
  for right particles. The red line is the typical path, corresponding to the classical trajectory.
  The parameter $\c=\cos(\theta)$ with $\theta=\pi/8$ here corresponds to $m\simeq .92\emm$.
  (Center) Details in 3D of the evolution of the single-particle Gauss-packet limited to dimension
  $64$ and for 20 time-steps. (Right) A ``double-slit'' particle state
  $|\Psi\>=\frac{1}{2}(|\uparrow_n^+\>+|\uparrow_n^-\> +|\uparrow_{-n}^+\>+|\uparrow_{-n}^-\>)$ for
  $n=10$, 80 time-steps, and $\theta=\pi/10$.  }\label{f:gausspak}
\end{figure}

In this section we come back to the $D=1$ Dirac automaton. The number of up-qubits
$S=\sum_n\sigma_n^z$ is a constant of motion. The observable algebra thus can be decomposed as
${\cal A}=\bigoplus_{S=0}^\infty{\cal A}_S$, and one can consider the evolution within each single
subalgebra separately.  A basis for the single-particle Fock sector is given by the orthonormal set
\begin{equation}
\{|\psi_n^\alpha\>\}_{\alpha=\pm,n\in\mathbb{Z}},\quad |\psi_n^\alpha\>:=\psi_n^\alpha{}^\dag|\Omega\>,
\end{equation}
or, in tensor notation
\begin{equation}\label{tensor-s}
|\psi_n^\alpha\>=|n\>\otimes|\alpha\>.
\end{equation}
Using the invariance of the vacuum $U|\Omega\>=|\Omega\>$, and the identity $U|\vpsi_n\>
=U\vpsi_n^\dag U^\dag
|\Omega\>=\vpsi_n^\dag\MU|\Omega\>=\MU|\vpsi_n\>$, we get
\def\P{\begin{matrix}\ldots&\ldots\\\ldots&\ldots\end{matrix}}
\def\L{\begin{matrix}0&-i\c\\-i\c&0\end{matrix}} \def\Z{\begin{matrix}\s &0 \\0& 0\end{matrix}}
\def\M{\begin{matrix}0 &0 \\0&\s\end{matrix}} \def\z{\begin{matrix}0&0 \\0&0\end{matrix}}
\begin{equation}\label{UABC}
U|\vpsi_n\>=\left[\Z\right]|\vpsi_{n-1}\>+\left[\L\right]|\vpsi_{n}\>+\left[\M\right]|\vpsi_{n+1}\>,
\end{equation}
or, in tensor notation,
\begin{equation}
U=\s(e_-\otimes|\uparrow\>\<\uparrow|+e_+\otimes|\downarrow\>\<\downarrow|)
-i\c I\otimes \sigma^1, 
\end{equation}
where $e_\pm:=\sum_n|n\pm 1\>\<n|$. The evolution of single-particle states
\begin{equation}
|\Psi^{(1)}\>=\sum_{ns}\Psi_{n\alpha}^{(1)}|\psi_n^\alpha\>,\quad |\psi_n^\alpha\>=|\!\uparrow_n^\alpha\>=
\begin{cases}
|\!\uparrow_{2n}\>, & \alpha=+\\
|\!\uparrow_{2n+1}\>, & \alpha=-
\end{cases}
\end{equation}
is a unitary band-matrix with matrix blocks given by Eq. (\ref{UABC}). In Fig. \ref{f:gausspak} as
an example we give the computer simulation of a Gaussian single-particle state of the form
\begin{equation}\label{Gauss-pa}
|\Psi\>=N^{-\frac{1}{2}}\sum_n e^{i\frac{2\pi n}{k}-\frac{(n-n_0)^2}{2\Delta^2}}  
(|\psi_n^+\>\pm |\psi_n^-\>),\quad
+:\text{ particle},\; -:\text{ antiparticle}.
\end{equation} 
In the same Fig. \ref{f:gausspak} we also see a {\em double-slit} particle localized state evolving.
It is noticeable how the Gaussian packet spreads very little compared to localized states.  One can
see that the following states are invariant
\begin{equation}\label{stablevectors}
\Psi^{(1)}_{n\alpha}(\phi)\propto
\begin{bmatrix}
\s\, \sin\phi+\alpha\xi\\ \c
\end{bmatrix}e^{in\phi},\qquad \xi=\sqrt{1-\s^2\cos^2\phi}.
\end{equation}
Notice that such states are not normalizable for unbounded LQCA. For small mass (and or small
``momentum'' $\sin\phi=\epsilon$ ) the invariant states approach left and right-handed states,
whereas at the Planck mass limit they become particle/antiparticle states.

In the single-particle sector one can define the position and momentum operators
\begin{equation}\label{PQ}
P:=\sum_\alpha P^\alpha,\; X:=\sum_\alpha X^\alpha,\quad
P^\alpha=-i\hbar\sum_n|\psi_n^\alpha\>\widehat\partial_x\<\psi_n^\alpha|,\;
X^\alpha=2a\sum_n|\psi_n^\alpha\>n\<\psi_n^\alpha|.
\end{equation}
and check the commutation relations $[X^\alpha,P^\beta]=i\hbar I_{\alpha}\delta_{\alpha\beta}$, and
$[X,P]=i\hbar I_1$.
For the invariant states (eigenstates of the momentum), one has
$\vec P|\vpsi\>=\frac{\hbar}{\ell}\sin\phi|\vpsi\>$.

The general form of a two-particle state is ($\psi^+_n=\varphi_{2n}$, $\psi^-_n=\varphi_{2n+1}$)
\begin{equation}\label{2part}
|\Psi^{(2)}\>=\sum_{ij}\Psi^{(2)}_{ij}\varphi_j^\dag\varphi_i^\dag|\Omega\>,\qquad
\Psi_{ij}^{(2)}=-\Psi_{ji}^{(2)}.
\end{equation}
Clearly the (normalized) antisymmetrized Kronecker product of two single-particle vectors is a
two-particle state, and it is also easy to prove that its unitary evolution is just the
antisymmetrized Kronecker product of the unitarily evolved single-particle states.

\begin{figure}[b]
\includegraphics[width=.24\textwidth]{\figdir 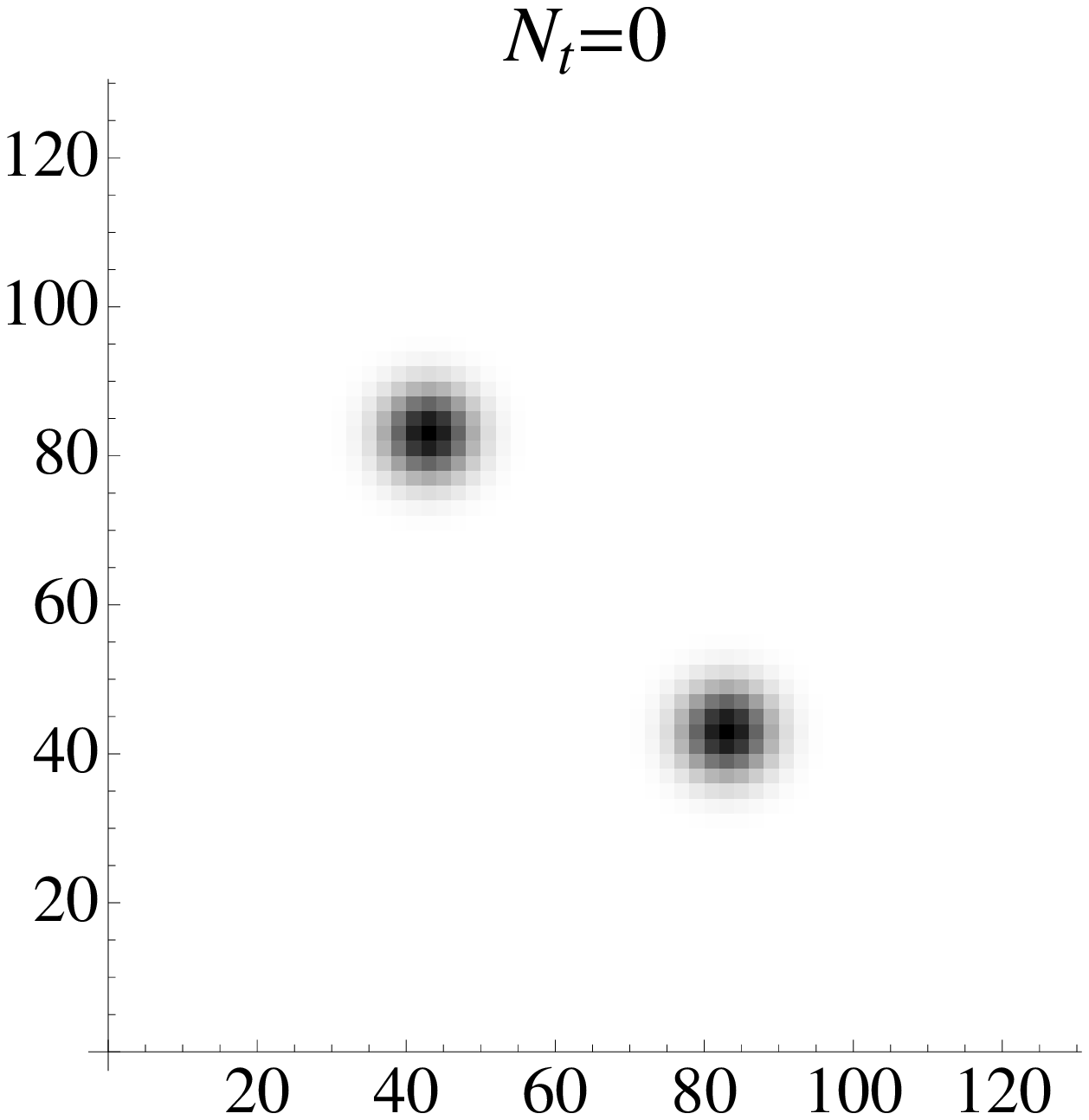} 
\includegraphics[width=.24\textwidth]{\figdir 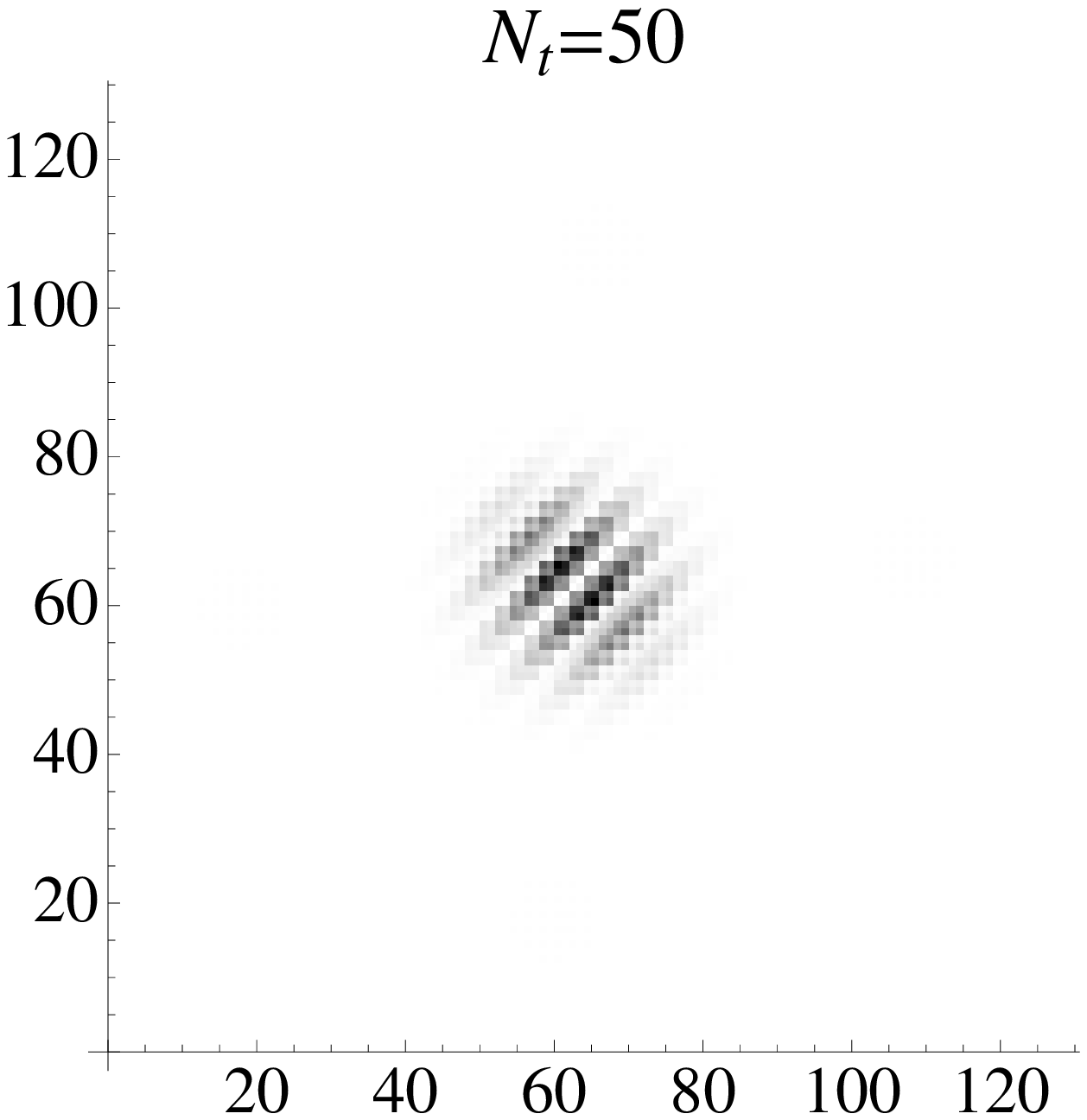} 
\includegraphics[width=.24\textwidth]{\figdir 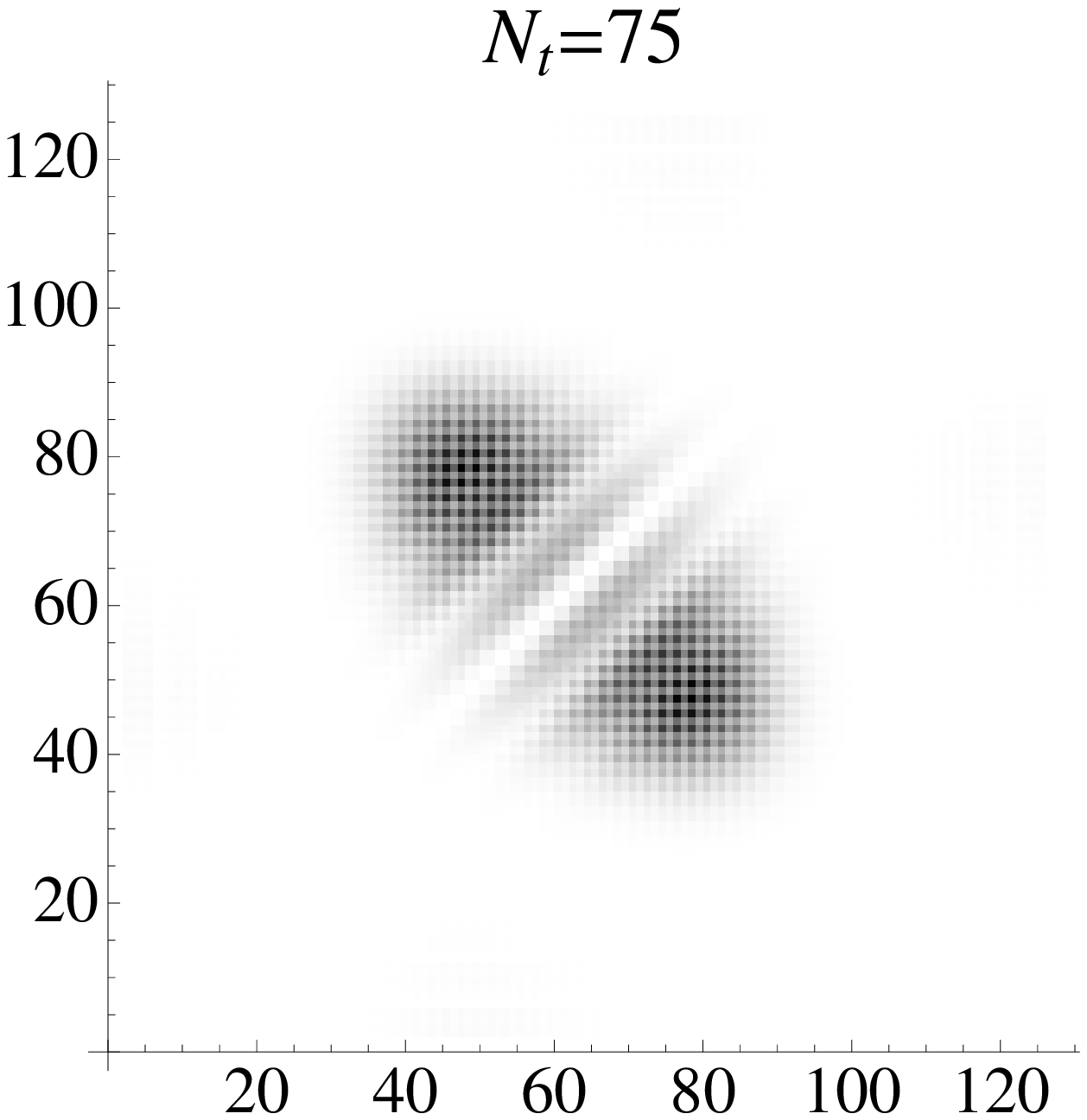} 
\includegraphics[width=.24\textwidth]{\figdir 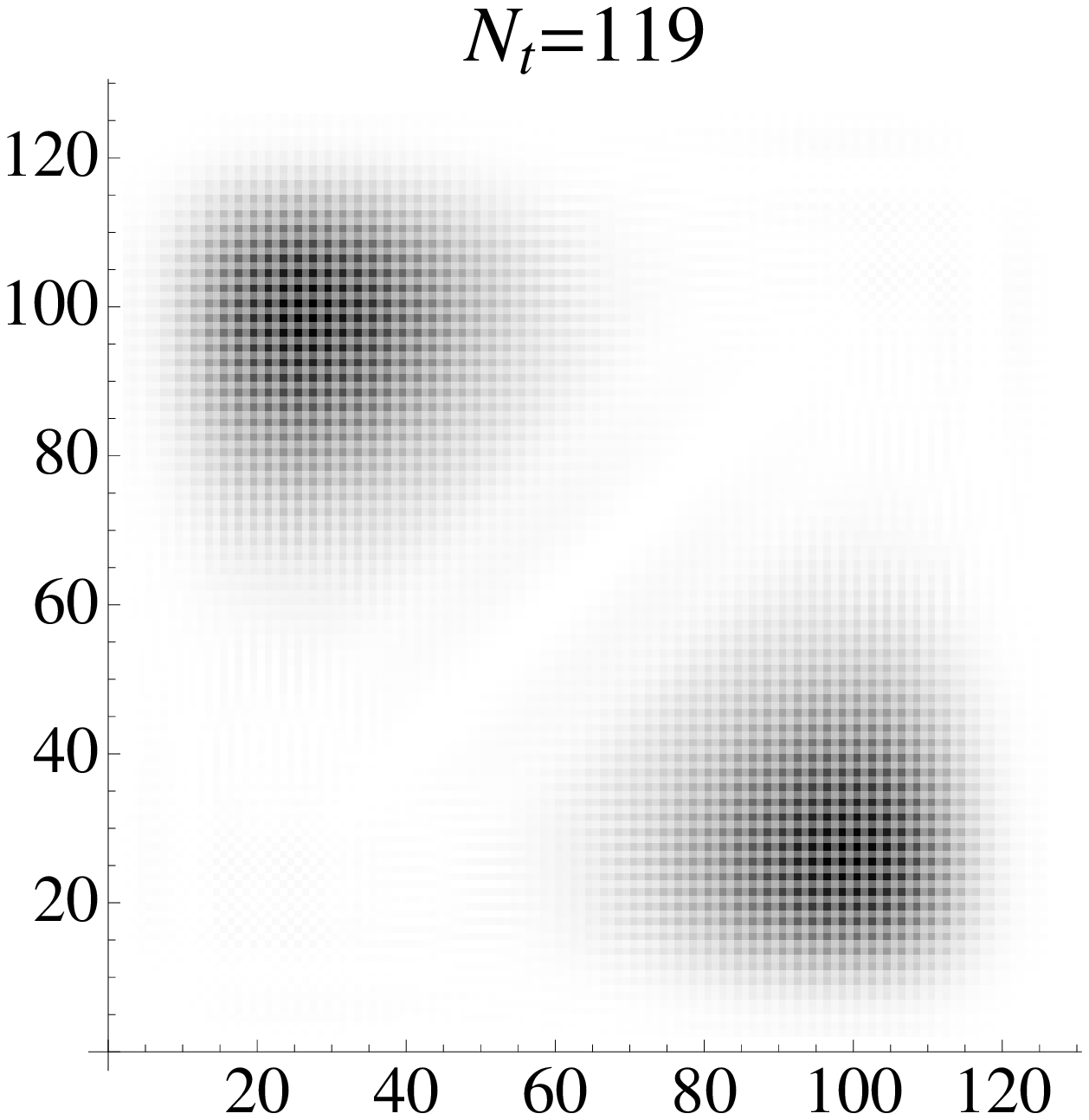} 
\caption{Evolution at varying time-steps $N_t$ of a two-particle state made upon anti-symmetrizing
  the two Gaussian packets in collision with $x_0=\pm 10$, $\Delta=2$, $k=\pm 8$, $128$ qubits.
  Here $m\simeq .92\emm$ as in Fig.  \ref{f:gausspak}.  The plot represents the matrix
  $|\Psi_{nm}^{(2)}|^2$ in Eq.  (\ref{2part}), with the vertical and the horizontal axes denoting
  the one-dimensional coordinate of the two particles. The symmetry along the diagonal is the result
  of indistinguishability.}\label{f:2Gauss}
\end{figure}

Finally, regarding the mentioned {\em classicalization} procedure, the derivation of the classical
field theory corresponding to the quantum cellular automaton proceeds as follows. We build up the
fields from qubits via the Jordan Wigner transformation, deriving the classical Hamiltonian via the
procedure in Subsection \ref{semerg}. Then, the classical particle emerges from Gaussian
single-particle states, with position and momentum given by the expectations of the operators in Eq.
(\ref{PQ}). The classical trajectory is the ``typical path'' along the quantum network, namely the
path with maximum probability (see e.~g. Fig. \ref{f:gausspak}).
\section{Conclusions and open problems}\label{s:conclu}
We have explored the quantum cellular automaton introduced in Ref. \cite{later} to implement the
Dirac equation in (1+1) dimension as a quantum computation. The automaton can be regarded as a {\em
  quantum ab-initio field theory}, i.~e. made with QT only with the addition of the requirements of
interaction-locality, translational invariance, and reversibility. From the first computer
simulations we have seen that at large scales the QCA mimics the corresponding QFT theory, whereas
interesting departures arise at the Planck scale, for high mass and momentum. As mentioned, one of
the most urgent open problems is now to devise an asymptotic method to evaluate the leading
corrections for large number of time-steps and states smoothly varying over large number of sites,
in order to derive the phenomenological modifications occurring at the Planck scale, of the kind of
e.~g. of Ref. \cite{amelino}. This task is made difficult by the simple fact that discrete automata
generally exhibit chaos even in the quantum case (see e.~g. Ref.  \cite{kicktop}): here, however, a
simplification is provided by the field-linearity of the automaton.

We have seen how the Feynman problem of simulating the field automaton by a local-matrix quantum
automaton is solved for $D=2$, and similarly can be solved for $D=3$. Additional problems, however,
must be addressed for space dimension $D>1$. For example, we still must write down explicitly a
Dirac automaton for $D>1$ (notice that generally there exist many automata mimicking the same linear
field equation). An instance of such automaton is the one provided by Bialynicki-Birula
\cite{BialynickiBirula}, which is proved to approach the Dirac equation in the continuum limit
$\ell\to0$ (via Trotter's formula), but it is unknown how it would behave in the large-scale limit
of infinitely many steps and smooth delocalized states.  A more interesting possibility is to
consider the emergent Hamiltonian for the automaton version of the QFT theory, and write the
Hamiltonian as a difference of the unitary matrix of the automaton and its adjoint, as in Eq.
(\ref{Schr}).  The {\em Weyl tiling} issue for $D>1$, on the other hand, clearly pose no problem for
the automaton realization, since it would persist in the continuum limit, but is contradicted by the
continuum limit of the Bialynicki-Birula automata. This shows that the {\em quantum} nature of the
causal network plays a crucial role in having Minkowski space-time as emergent from the discrete
geometry of pure topology. It would be interesting to see explicitly how the quantum superposition
of different paths on the network restores the metrical isotropy.

\subsection*{Acknowledgments} 
I thank Norman Margolus, Reinhard Werner, and Seth Lloyd for interesting discussions and
suggestions, and Gianluca Introzzi for his careful reading of the manuscript.


\begin{thebibliography}{0}
\bibitem{cdpax} G. Chiribella, G. M. D'Ariano, P. Perinotti, {\em Informational derivation of
    quantum theory}, Phys. Rev. A {\bf 84} 012311 (2011) [Free download from the {\em Viewpoint} in
  Ref. \cite{viewpoint}.
\bibitem{viewpoint} \v{C}aslav Brukner, {\em Questioning the rules of the game}, Physics {\bf 4} 55 (2011),
{\em Viewpoint} of Ref. \cite{cdpax}.
\bibitem{my2005} G. M. D'Ariano, \emph{On the missing axiom of quantum mechanics},
  Quantum Theory, Reconsideration of Foundations - 3, V\H{a}xj\H{o}, Sweden, 6-11 June 2005
  (Melville, New York) (G.~Adenier, A.~Y. Khrennikov, and T.~M. Nieuwenhuizen, eds.), (AIP,
  Melville, New York 2006) pp.~114.
\bibitem{myCUP2009} G. M. D'Ariano, in \emph{Philosophy of quantum information and entanglement},
  A.~Bokulich and G.~Jaeger eds., (Cambridge University Press, Cambridge UK, 2010). Also arXive
  0807.4383.
\bibitem{first} M. D'Ariano, {\em On the "principle of the quantumness", the quantumness of Relativity,
  and the computational grand-unification}, in AIP CP {\bf 1232} Quantum Theory: Reconsideration of Foundations,
 5, edited by A.~Y. Khrennikov (AIP, Melville, New York, 2010), pag 3 (also arXiv:1001.1088).
\bibitem{amelino} G. Amelino-Camelia, {\em Relativity in space-times with short-distance structure
    governed by an observer-independent (Planckian) length scale}, Int. J. Mod Phys. {\bf D11} 35
  (2002) [also arXiv:gr-qc/0012051]
\bibitem{later} G. M. D'Ariano, {\em The Quantum Field as a Quantum Computer}, arXiv (2010)
\bibitem{tosini} G. M. D'Ariano and A. Tosini, {\em Emergence of Space-Time from Topologically
    Homogeneous Causal Networks}, Submitted to Studies in History and Philosophy of Modern Physics
  special Issue on Emergent Space-Time, arXiv:1109.0118 [see also arXiv:1008.4805 (2010) for a
  preliminary account]
\bibitem{weyl} H. Weyl, {\em Philosophy of Mathematics and Natural Sciences}, (Princeton University
  Press, Princeton 1949)
\bibitem{tobias} T. Fritz, {\em Velocity Polytopes of Periodic Graphs}, arXiv:1109.1963 
\bibitem{FQXi} G. M. D'Ariano, {\em A Quantum-digital Universe}, FQXi Essay Contest {\em Is Reality Digital or
  Analog?} (2011) [{\tt http://fqxi.org/community/essay/winners/2011.1}]
\bibitem{vaxjo2010} G. M. D'Ariano, {\em Physics as Information Processing}, in AIP Conf. Proc. {\bf
    1327} Advances in Quantum Theory, edited by  G. Jaeger, A. Khrennikov, M. Schlosshauer, and
  G. Weihs (AIP, Melville, New York, 2011) pag. 7
\bibitem{QCM2010} G. M. D'Ariano, {\em Physics as quantum information processing}
QCMC 2010 Brisbane, Australia (in press) (also arXiv:1012.2597)
\bibitem{yepez} J. Yepez, {\em Relativistic Path Integral as a Lattice-based Quantum Algorithm},
Q. Inf. Proc. {\bf 4} 471 (2006)
\bibitem{Feynman} R. P. Feynman, {\em Simulating Physics with Computers}, Int. J. Th. Phys,
{\bf 21} 467 (1982)
\bibitem{werner} P. Arrighi, V. Nesme, and R. Werner, {\em Unitarity plus causality implies
    localizability}, arXiv:0711.3975
\bibitem{Margolus} T. Toffoli and M. Margolus, {\em Invertible Cellular automata: a review}, Physica
  D {\bf 45} 229 (1990)
\bibitem{Normbook} T. Toffoli and N. Margolus, {\em Cellular Automata Machines: A New Environment
    for Modeling} (MIT Press, Cambridge, USA 1987).
\bibitem{Deutsch} D. Deutsch, {\em Quantum theory, the Church-Turing principle, and the universal
    quantum computer}, Proc. R. Soc. Lond. {\bf 400} 97 (1985)
\bibitem{JW} P. Jordan and E. Wigner, {\em About the Pauli exclusion principle}, Z. Phys. {\bf 47} 631 (1928)
\bibitem{feynmann} R. P. Feymann, {\em Simulating Physics with Computers}, Int. J. Theor. Phys. {\bf
    21} 467 (1982)
\bibitem{Fradkin} E. Fradkin, {\em Jordan Wigner transformation for quantum-spin systems in two
    dimensions and fractional statistics} Phys. Rev. Lett. {\bf 63} 322 (1989) 
\bibitem{Eliezer} D. Eliezer and G. W. Semenoff, {\em Intersections Forms and the Geometry of
      Lattice Chern-Simons Theory}, Phys. Lett. B {\bf 286} 118 (1992)
\bibitem{Huerta} L. Huerta and J. Zanelli {\em Bose-fermi transformation in
    three-dimensional space}, Phys. Rev. Lett. {\bf 71} 3622 (1993)
\bibitem{Zee} Q. Zee, {\em Quantum Field Theory in a Nutshell} (Princeton Scottsdale 2010)
\bibitem{kicktop} G. M. D'Ariano, L. R. Evangelista, M. Saraceno
{\em Classical and Quantum Structures in the Kicked Top Model}, Phys. Rev. A {\bf 45} 3646 (1992)
\bibitem{BialynickiBirula} I Bialynicki-Birula, {\em Weyl, Dirac, and Maxwell equations on a lattice as
    unitary cellular automata}, Phys. Rev. D {\bf 49} 6920 (1994)
\bibitem{cirvar} F. Verstraete and J. I Cirac, {\em Mapping local Hamiltonians of fermions to local
    Hamiltonians of spins}, J. Stat. Mech. {\bf 09} 12  (2005)
\end{thebibliography}
\end{document}